\documentclass[11pt]{article}
\usepackage{amsmath,amssymb,color,graphics,epsfig}

\usepackage{amsfonts}

\newcommand{\be}{\begin{equation}}
\newcommand{\ee}{\end{equation}}
\newcommand{\bea}{\setlength\arraycolsep{2pt} \begin{eqnarray}}
\newcommand{\eea}{\end{eqnarray}}
\newcommand{\nn}{\nonumber}

\def\ft#1#2{{\textstyle{\frac{\scriptstyle #1}{\scriptstyle #2} } }}
\def\fft#1#2{{\frac{#1}{#2}}}

\def\0{{\sst{(0)}}}
\def\1{{\sst{(1)}}}
\def\2{{\sst{(2)}}}
\def\3{{\sst{(3)}}}
\def\4{{\sst{(4)}}}
\def\5{{\sst{(5)}}}
\def\6{{\sst{(6)}}}
\def\7{{\sst{(7)}}}
\def\8{{\sst{(8)}}}
\def\sst#1{{\scriptscriptstyle #1}}

\begin{document}

\begin{flushright}
\end{flushright}

\vspace{25pt}
\begin{center}
{\large {\bf Global Structure of Exact Scalar Hairy Dynamical Black Holes}}

\vspace{10pt}
Zhong-Ying Fan$^{1}$\,, Bin Chen$^{1,2,3}$ and H. L\"u$^{4}$

\vspace{10pt}
{\it $^{1}$Center for High Energy Physics, Peking University, No.5 Yiheyuan Rd,\\}
{\it  Beijing 100871, P. R. China\\}
\smallskip
{\it $^{2}$Department of Physics and State Key Laboratory of Nuclear Physics and Technology,\\}
{\it Peking University, No.5 Yiheyuan Rd, Beijing 100871, P.R. China\\}
\smallskip
{\it $^{3}$Collaborative Innovation Center of Quantum Matter, No.5 Yiheyuan Rd,\\}
{\it  Beijing 100871, P. R. China\\}
\smallskip
{\it $^{4}$Center for Advanced Quantum Studies, Department of Physics,\\ }
{\it Beijing Normal University, Beijing 100875, P. R. China}

\vspace{40pt}

\underline{ABSTRACT}
\end{center}
We study the global structure of some exact scalar hairy dynamical black holes which were constructed in Einstein gravity either minimally or non-minimally coupled
to a scalar field. We find that both the apparent horizon and the local event horizon (measured in luminosity coordinate) monotonically increase with the advanced time
as well as the Vaidya mass. At late advanced times, the apparent horizon approaches the event horizon and gradually becomes future outer. Correspondingly, the space-time
arrives at stationary black hole states with the relaxation time inversely proportional to the $1/(n-1)$ power of the final black hole mass, where $n$ is the space-time dimension. These results strongly support the solutions describing the formation of black holes with scalar hair. We also
obtain new charged dynamical solutions in the non-minimal theory by introducing an Maxwell field which is non-minimally coupled to the scalar.
 The presence of the electric charge strongly modifies the dynamical evolution of the space-time.

\vfill {\footnotesize Emails: fanzhy@pku.edu.cn,\quad bchen01@pku.edu.cn,\ \ \ mrhonglu@gmail.com}

\thispagestyle{empty}

\pagebreak

\tableofcontents
\addtocontents{toc}{\protect\setcounter{tocdepth}{2}}




\section{Introduction}

 The no-hair theorems in General Relativity exclude the existence of hairy black holes in asymptotically flat space-times for a variety of theories. However,
it turns out that the no-hair theorems are easily evaded. People have found many counter examples in recent years. For instance, many scalar hairy black holes that are asymptotic
to Minkowski space-times have been analytically constructed in Einstein gravity minimally coupled to a scalar field \cite{Anabalon:2013qua,Anabalon:2012dw,Gonzalez:2013aca,Feng:2013tza,Fan:2015oca}. Rotating black holes with scalar hair were also numerically studied in \cite{Herdeiro:2014goa}.
Black holes with vector hair \cite{Geng:2015kvs} and with Yang-Mills hair \cite{Bizon:1990sr,Kleihaus:2000kg,Kleihaus:2002ee,Meessen:2008kb,Fan:2014ixa} have also
been found in literature. By numerical analysis, it was established that non-Schwarzschild black holes do exist in higher derivative gravity in four
 dimensions \cite{Lu:2015cqa,Lu:2015psa}.

For asymptotical (anti-)de Sitter space-times, the condition for the no-hair theorems was much relaxed. There are large classes of scalar hairy black holes having
been constructed in Einstein gravity either minimally or non-minimally coupled to a scalar field \cite{Anabalon:2012dw,Gonzalez:2013aca,Feng:2013tza,Fan:2015oca,Henneaux:2002wm,Martinez:2004nb,Anabalon:2013sra,Acena:2013jya,Zloshchastiev:2004ny,Fan:2015tua}.
More interestingly, some of these theories even admit exact dynamical black holes solutions with scalar hair \cite{Fan:2015tua,Zhang:2014sta,Zhang:2014dfa,Lu:2014eta,Xu:2014xqa,Ayon-Beato:2015ada,Fan:2015ykb}. The solutions provide explicit and
analytical examples for the formation of scalar hairy black holes. In particular, one class of solutions analytically shows how the linearly stable AdS vacua
undergoes non-linear instability and spontaneously evolves into stationary black holes states \cite{Fan:2015tua,Fan:2015ykb}. The solutions that are asymptotic to AdS
space-times also have potential applications in the AdS/CFT correspondence.


The purpose of current paper is to gain an even deeper understanding of the scalar hairy dynamical solutions which were found in certain minimally and non-minimally
coupled Einstein-scalar gravity by studying their global properties. We focus on discussing the dynamical evolution of the apparent horizon which was studied as a
function of the advanced time. We find that the apparent horizon (measured in luminosity coordinate) grows monotonically with the advanced time and approaches the event horizon at the future
infinity. At late times of the evolution, the apparent horizon smoothly evolves and becomes future outer gradually. In addition, it is also instructive to study the dynamical evolution of the event horizon for our solutions.
Unfortunately, the event horizon is in general hard to be established in dynamical space-times, though it is globally well defined. Instead, we adopt an effective notion ``local event horizon". We find that it
covers the apparent horizon in the whole dynamical process and also approaches the event horizon at the future infinity. These results provide strong evidence to support
the dynamical solutions describing black holes formation.

Moreover, we generalize the non-minimally coupled Einstein-scalar gravity by introducing an additional Maxwell field which is non-minimally coupled to the scalar.
We obtain new charged dynamical solutions for proper gauge coupling functions. The dynamical evolution of the space-time turns out to be strongly dependent on the
electric charges. The global property of the charged solutions differs significantly from the pure neutral solutions.

This paper is organized as follows. In section 2, we present some preliminaries for studying the global properties of dynamical solutions. In section 3,
we study two explicit examples: the dynamical solutions that are constructed in certain minimally and non-minimally coupled Einstein-scalar gravity.
In section 4, we obtain more charged dynamical solutions in the non-minimal theory and discuss their global properties.
We conclude this paper in section 5.

\section{Global properties of dynamical solutions}

In this section, we give the preliminaries for studying the global structure of dynamical solutions of the type
\be
ds^2=-Hdu^2 + 2 h du dr + \rho^2 d\Omega_{n-2}^2\,,
\ee
where $H, h$ and $\rho$ are all functions of the Eddington-Finkelestein-type coordinates $r$ and $u$ and $d\Omega_{n-2}$ is the $(n-2)$ dimensional space with
spherical/torus/hyperbolic symmetries.

\subsection{Local event horizon}

The event horizon of a static black hole can be easily established corresponding to a null Killing vector with non-negative surface gravity.
The situation for a dynamical solution is much more subtle.  The location of the event horizon cannot be solved in general even for an exact black hole solution.
We shall adopt a local definition for the event horizon, which can be found in many standard textbooks and some early literature such as \cite{Zhao:1992ad,Zhao:1995jr}. We call it ``local event horizon".
It is defined by a null hypersurface which preserves the isometry of the space-time. That is, if the null hypersurface is parameterized by:
\be F(r\,,u)=0\,,\ee
with $r=r(u)$, it satisfies
\be g^{\mu\nu}\fft{\partial F}{\partial x^{\mu}}\fft{\partial F}{\partial x^{\nu}}=0\,. \ee
This gives
\be H\fft{\partial F}{\partial r}+2h\fft{\partial F}{\partial u}=0 \,.\ee
On the other hand, we have
\be  0\equiv \fft{d F}{du}=\fft{\partial F}{\partial u}+\fft{\partial F}{\partial r}\fft{dr}{du} \,.\ee
Combining these two equations, we find
\be \fft{dr}{du}=\fft{H}{2h} \,.\label{eh}\ee
 It follows that the local event horizon coincides with the ``true" event horizon in the stationary limit. However, we should point out that it is not clear what
the precise relation is between the local event horizon and the event horizon in dynamical space-times. In this paper, we simply take it to be an effective conception for describing black holes
formation. Indeed, we find that it enjoys some reasonable aspects for our solutions. For example, the local event horizon always encloses the apparent horizon in
the dynamical process and approaches the event horizon at the future infinity.

\subsection{The apparent horizon}
For the dynamical solutions with planar topology describing gravitational collapse, a sufficient condition is the apparent horizon should be future outer at late advanced times \cite{Wang:2003bt}. We shall first give a general discussion on this topic.
The tangent vector of radial null geodesic congruences can be given by:
\be k_{\epsilon}^a\equiv(\ft{\partial}{\partial \xi_{\epsilon}})^a=\epsilon\Big((\ft{\partial}{\partial u})^a+\ft{H}{2h}(\ft{\partial}{\partial r})^a \Big)\,, \ee
where $\epsilon=-1$ for ingoing null geodesic congruences and $\epsilon=+1$ for outgoing null geodesic congruences respectively.
It is straightforward to verify that $k_{\epsilon}^a$ is null and it satisfies the null geodesic equations of motions, namely
\be k_{\epsilon}^b\triangledown_b k_{\epsilon}^a=\lambda_{\epsilon} k_{\epsilon}^a\,,\qquad \lambda_{\epsilon}=\fft{\epsilon}{2h}(2h_u+H_r) \,,\ee
where $H_r\equiv\fft{\partial H}{\partial r},\ h_u\equiv \fft{\partial h}{\partial u}$.
Given the tangent vector, it is easy to calculate the expansion
\be \theta_{\epsilon}=g_{ab}\triangledown^bk_{\epsilon}^a-\lambda_{\epsilon}=\ft{\epsilon(n-2)}{2}\Big(\fft{H\rho_r}{h\rho}+\fft{2\rho_u}{\rho}\Big) \,,\ee
where $\rho_r\,,\rho_u$ denotes the derivatives of $\rho$ with respect to $r$ and $u$ respectively. The location of the apparent horizon is defined by
$\theta_\epsilon=0$. In the luminosity coordinate
$\rho\equiv r$, the expansion was simplified to be $\theta_\epsilon=\ft{\epsilon(n-2)H}{2h r}$. It is immediately seen that the expansion of the ingoing null geodesic congruences $\theta_-$
is negative\footnote{The metric function $h$ should be positive definite outside the apparent horizon since $u$ was interpreted as the advanced
time coordinate.} outside the apparent horizon($H(r,u)>0$), which implies that the apparent horizon is future. This is independent of the coordinate system
and is valid for more general $\rho$ coordinate.

In order to show the apparent horizon is outer, we need compute the Lie derive of the expansion of the ingoing null geodesic congruences along the tangent
 vector of the outgoing null geodesic congruences
\bea\label{lie} \mathcal{L}_{k_+}\theta_{-}&=&k_+^b\triangledown_b\theta_-=k_+^r\fft{\partial \theta_-}{\partial r}+k_+^u\fft{\partial \theta_-}{\partial u}\\
 &=&\fft{(n-2)}{4\rho^2 h^3}\Big( 4h^3(\rho_u^2-\rho \rho_{uu})-2h^2\rho \rho_r H_u \nn\\
  &&+\big(4 h(\rho_r\rho_u-\rho\rho_{ru})+\rho\rho_r ( 2 h_u - H_r) \big)h H \nn\\
  && +\big(h\rho_r^2+\rho(h_r\rho_r-h\rho_{rr})\big)H^2 \Big) \,.\nn\eea
If this is also negative outside the apparent horizon, the apparent horizon is called outer. This is valid for planar black holes\footnote{It is easy to check that this condition does not hold any longer for spherical/hyperbolic black holes.}.
For the solutions considered in this paper, we will show that the apparent horizon approaches
the event horizon and smoothly evolves into a future outer one at the late times of the evolution.

\section{Explicit examples}

\subsection{The non-minimal example}
The first example we consider is non-minimally coupled Einstein-scalar gravity in general $n$ dimensions \cite{Fan:2015tua}. The Lagrangian is given by:
\be
\mathcal{L}_n=\sqrt{-g}\Big(\kappa_0 R-\ft12\xi\phi^2 R
-\ft12 (\partial \phi)^2-V(\phi)\Big) \label{gen-nonmin-lag}\,,
\ee
where $\kappa_0$ is the bared gravitational coupling constant and
$\xi$ is a constant that characterizes the coupling strength between the scalar $\phi$ and the curvature. The covariant equations of motion are
\be
E_{\mu\nu}\equiv\kappa_0 G_{\mu\nu}-T_{\mu\nu}^{\rm (min)}-T_{\mu\nu}^{\rm (non)}\,,\qquad
\Box\phi = \xi\phi\,R + \fft{dV}{d\phi}\,,\label{non-min-geneom}
\ee
where $ G_{\mu\nu}=R_{\mu\nu}-\fft 12 R g_{\mu\nu}$ is the Einstein tensor and
\bea
T_{\mu\nu}^{\rm (min)}&=&\ft 12\partial_\mu\phi \partial_\nu \phi-\ft 12 g_{\mu\nu}\Big(\ft 12(\partial \phi)^2+V(\phi) \Big)\,,\nn\\
T_{\mu\nu}^{\rm (non)}&=&\ft12\xi(\phi^2 G_{\mu\nu}+g_{\mu\nu}\square \phi^2-\nabla_\mu \nabla_\nu \phi^2)\,.
\eea
Without loss of generality, we let
\be
\kappa_0=\fft{\xi}{2}\,,
\ee
in this subsection. The effective gravitational coupling constant depends on $\phi$, given by
\be \kappa(\phi)=\kappa_0-\ft 12 \xi \phi^2=\ft 12 \xi(1-\phi^2) \,.\ee
We shall require $\phi<1$ to cancel ghost-like graviton modes.
For generic $\xi$, there are large classes of static hairy planar black holes which were reported in \cite{Fan:2015tua}. In particular,
When $\xi$ takes the value
\be
\xi=\frac{n-2}{4(n-1)}\,,
\label{xi}\ee
the non-minimal theory admits exact dynamic solutions with the potential given by
\be V=-\ft18(n-2)^2 \Big(g^2 + \alpha \phi^{\fft{2(n-1)}{n-2}}\big(\ft{1}{1-\phi^2}-
{}_2F_1[1,\ft{n-1}{n-2};\ft{2n-3}{n-2};\phi^2]\big)\Big)\,.\label{nonpotential}\ee
The scalar potential has a stationary point $\phi=0$ and its small $\phi$ expansion is given by
\be
V=-\ft18(n-2)^2 g^2- \fft{(n-2)^3\alpha}{8(2n-3)} \phi^{4 + \fft{2}{n-2}} -
\fft{(n-2)^3\alpha}{4(3n-5)} \phi^{6 + \fft{2}{n-2}} + \cdots\,.
\ee
It follows that the scalar has vanishing mass square, which is above the Breitenlohner-Freedman (BF) bound\footnote{Due to the non-minimal coupling, the effective mass of the scalar becomes position dependent, given by $m_{\mathrm{eff}}^2=m^2+\xi R$. Consequently, the BF bound is shifted by the non-minimal coupling constant $m^2_{BF}=-\fft{1}{4} (n-1)\Big((1-4\xi)n-1\Big)g^2$.} because $m_{\mathrm{BF}}^2=-\ft 14 g^2$ for the non-minimal coupling (\ref{xi}). This implies that the AdS vacua is linearly stable against perturbation.

In Eddington-Finkelstein-like coordinates, our dynamical solutions can be cast into the form of
\bea
ds^2 &=& - f du^2+2 du dr + r^2 dx^i dx^i\,,\qquad \phi=\Big(\fft{a}{r}\Big)^{\fft12(n-2)}\,,\cr
f&=& g^2 r^2 -\fft{\alpha a^{n-1}}{r^{n-3}}\,{}_2F_1[1,\ft{n-1}{n-2};
\ft{2n-3}{n-2}, \big(\ft{a}{r}\big)^{n-2}]\,,
\label{non-sol}\eea
where $a\equiv a(u)$ is the time dependent ``scalar charge". It satisfies a second order non-linear differential equation
\be
\fft{\ddot a}{a^2} - \fft{2\dot a^2}{a^3} + \fft{\tilde{\alpha}\dot a}{a}=0\,,\qquad \tilde{\alpha}=\ft 12(n-1)\alpha \,.
\label{non-aeq}\ee
Here a dot denotes the derivative with respect to $u$. This equation can be integrated. The resulting first order equation is
\be
\dot a + \tilde\alpha\, a^2\log \big(\fft{a}{q}\big)=0\,,\label{firstorder}
\ee
where $q$ is an integration constant. It is clear that both $a=0$ and $a=q$ are stationary points. The former corresponds to the pure AdS vacua whilst the latter corresponds
to the stable black holes. The first order equation can be solved exactly in terms of an exponential integral function
\be
{\rm Ei}\big(\log (\ft{q}{a})\big) = -\tilde\alpha q\,u\,.\label{aei}
\ee

\begin{figure}[ht]
\begin{center}
\includegraphics[width=230pt]{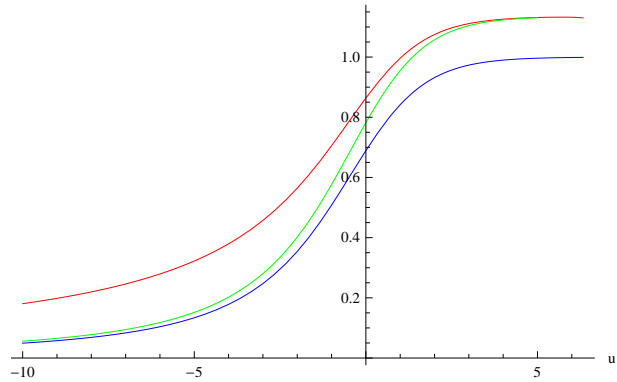}
\end{center}
\caption{{\it The plots for $a(u)$ (the blue line), the apparent horizon (the green line) and the local event horizon (the red line) for the $n=4$ dimension. At late advanced times $u\rightarrow \infty$, $a(u)$ approaches its equilibrium value $q$ and both horizons approach the event horizon at an exponential
 rate of $e^{-u/u_0}$, where $u_0=1/(\tilde{\alpha}q)$ is the characteristic relaxation time. Some constants $g^2\,,\tilde{\alpha}\,,q$ have been set to unity.}}
\label{fig1}\end{figure}
In Fig.\ref{fig1}, we plot $a(u)$ as a function of the advanced time $u$ (the blue line). The advanced time coordinate $u$ runs over ($-\infty\,,+\infty$). Correspondingly, the space-time evolves from pure AdS vacua with $a=0$ at the past infinity to stable black holes states with $a=q$ at the future infinity.
In addition, from the asymptotical behavior of the metric function $f$
\be f=g^2 r^2-\fft{\alpha a^{n-1}}{r^{n-3}}+\cdots \,,\ee
we can read off the effective Vaidya mass
\be M=\fft{(n-2)^2\alpha}{128(n-1)\pi}a^{n-1}\,,\label{bhmass}\ee
and deduce its first order derivative
\be \dot{M}=\fft{(n-1)(n-2)^2\alpha^2}{256\pi} a^n \log\big(\fft{q}{a}\big)\,.  \ee
It follows that $\dot{M}$ is positive for $a<q$, implying that the Vaidya mass monotonically increases with the advanced time.
\begin{figure}[ht]
\begin{center}
\includegraphics[width=200pt]{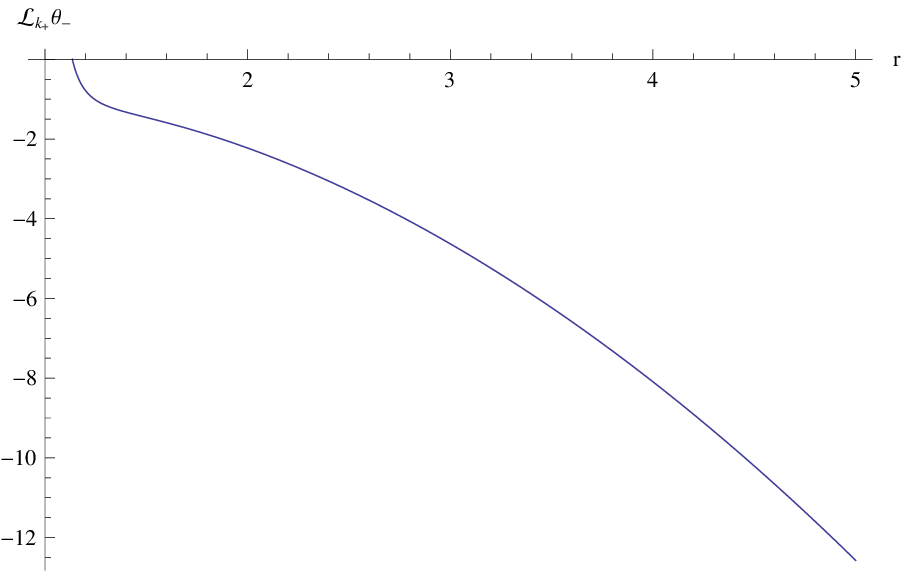}
\includegraphics[width=200pt]{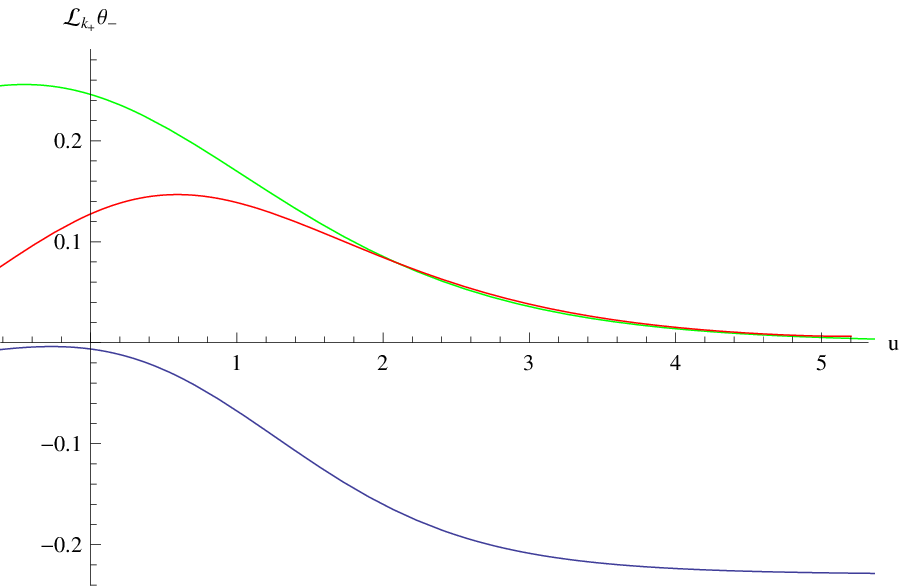}
\end{center}
\caption{{\it The plots for the Lie derivative $\mathcal{L}_{k_+}\theta_{-}$ for $n=4$ dimension. The left plot is given for static solutions and the Lie derivative is plotted as a function of $r$. In the right plot, the Lie derivative is plotted as a function of the advanced time $u$ for $r=r_{AH}$ (Green), $r=r_{EH}$ (Red) and $r=1.23$ (Blue), respectively. We have $r_{AH}<r_{EH}<1.23$. To have a nice presentation, we have properly scaled the Lie derivative on the apparent horizon as
$\mathcal{L}_{k_+}\theta_{-}\rightarrow \ft 15\mathcal{L}_{k_+}\theta_{-}$ and shifted it as $\mathcal{L}_{k_+}\theta_{-}\rightarrow \mathcal{L}_{k_+}\theta_{-}+0.715$ at $r=1.23$. Some constants $g^2\,,\tilde{\alpha}\,,q$ have been set to unity.  }}
\label{fig2}\end{figure}

 At the past infinity $u\rightarrow -\infty$, $a(u)$ behaves as $a\log a\sim -1/u$, indicating that the scalar only provides a weak source to perturb the space-time.
 However, the AdS vacuum is linearly stable against perturbation. It is some non-linear effects that pushes the space-time evolving into stationary black holes states.
 Hence, this solution provides an analytical example how the linearly stable AdS vacua undergoes non-linear instability and eventually settles
 down to stable black holes states.

In Fig.\ref{fig1}, we also plot the apparent horizon (the green line) and the local event horizon (the red line) during the black holes formation. It is interesting to note that the apparent horizon is always inside the local event horizon in the whole dynamic process and both of them approach the event horizon at the large $u$ region. In fact, at the future infinity $u\rightarrow +\infty$, the space-time approaches the static limit exponentially fast. We find
\bea &&a(u)=q\Big(1-c_0e^{-\tilde{\alpha}q u}+\ft{3c_0^2}{2}e^{-2\tilde{\alpha}q u}+\cdots \Big)\,,\nn\\
    &&M(u)=M_0\Big(1-(n-1)c_0e^{-\tilde{\alpha}q u}+\ft{(n^2-1)c_0^2}{2}e^{-2\tilde{\alpha}q u}+\cdots \Big) \,.\eea
Here $c_0$ is a positive integration constant which can be absorbed by constant shift of $u$ and $M_0$ is the static black hole mass, given by Eq.(\ref{bhmass})
with $a$ replaced by $q$. The apparent horizon and the local event horizon have analogous behaviors at late advanced times. Therefore, we can naturally define a
characteristic relaxation time $u_0=1/(\tilde{\alpha}q)\sim 1/M_0^{\fft{1}{n-1}}$. It turns out that the relaxation time becomes shorter for a bigger black hole
mass.

To end this subsection, we remark that our solutions have future outer apparent horizons. This is immediately seen from the behavior of the Lie derivative
$\mathcal{L}_{k_+}\theta_{-}$ (see Fig.\ref{fig2}). For the static solutions it vanishes on the apparent horizon and becomes negative outside the apparent horizon.
In the dynamical process, the Lie derivative continuously evolves and approaches the static limit at late advanced times. These results imply
 that the apparent horizon is future outer and the dynamical solutions describe the physical process of black holes formation.

\subsection{The minimal example}
The second example we consider is the dynamical black holes which were found in certain minimal coupled Einstein-scalar gravity in four dimensions
\cite{Zhang:2014sta}. The Lagrangian is given by
\bea
{\cal L} &=& \sqrt{-g} \Big(R - \ft12(\partial\phi)^2 - V(\phi)\Big),\cr
V&=& -2g^2 (\cosh\phi + 2) - 2\alpha (2\phi + \phi \cosh\phi - 3 \sinh\phi)\,.
\label{lagmini}
\eea
The scalar potential has a stationary point $\phi=0$ with small $\phi$ expansion
\be V=-6g^2 - g^2 \phi^2 - \ft1{12} g^2 \phi^4 -\ft{1}{30}\alpha^2 \phi^5+ \cdots \,.\ee
It follows that the scalar has a mass square $m^2=2g^2$ which is above the Breitenlohner-Freedman bound ($m_{BF}^2=-\ft 94 g^2$) in AdS space-time with $g^2>0$.
In fact, the potential arises from $\mathcal{N}=4\,,D=4$ gauged supergravity for vanishing $\alpha$ and $g^2>0$ \cite{Gates83}.

Using Eddington-Finekelstein-like coordinates, the dynamical solution reads
\bea
ds^2 &=& -f du^2 +2du dr + r(r+a) d\Omega_{2,k}^2\,,\qquad
\phi=\log{\Big(1 + \fft{a}{r}\Big)}\,,\cr
f&=&g^2r^2 + k -\ft12\alpha a^2-\dot{a} + (g^2-\alpha)a r+ \alpha r^2 \Big(1 + \fft{a}{r}\Big)\log\! \Big(1 + \fft{a}{r}\Big)\,,\label{minibh}
\eea
where $d\Omega_{2,k}$ is the $2$-space with constant curvature $k=0\,,\pm 1$. The time dependent ``scalar charge" $a\equiv a(u)$ satisfies
\be \ddot{a}+\alpha a \dot a=0\,,\ee
which can be solved immediately by
\be a=q \tanh{\Big(\ft 12\alpha q u \Big)} \,.\ee
It follows that $a=0$ at $u=0$. However, its first order derivative $\dot{a}=\ft 12 \alpha q^2 \mathrm{sech}{\Big(\ft 12\alpha q u\Big)}$ is non-vanishing at this point.
This results to subtle global structure around this point \cite{Zhang:2014sta}. When $u$ goes to zero, the space-time turns out to be pure AdS except the
singular point $R=0$, where $R$ is the
luminosity coordinate defined by $R=\sqrt{r(r+a)}$. In fact, the existence of this singularity strongly depends on the path to $(0\,,0)$ point on the $(u\,,R)$ plane. We find that for some paths such as $R\propto u^\delta$
with $\delta\leq 1/3$, the singularity disappears. However, all these trajectories are space-like, indicating that this singularity is non-traversable.


\begin{figure}[ht]
\begin{center}
\includegraphics[width=200pt]{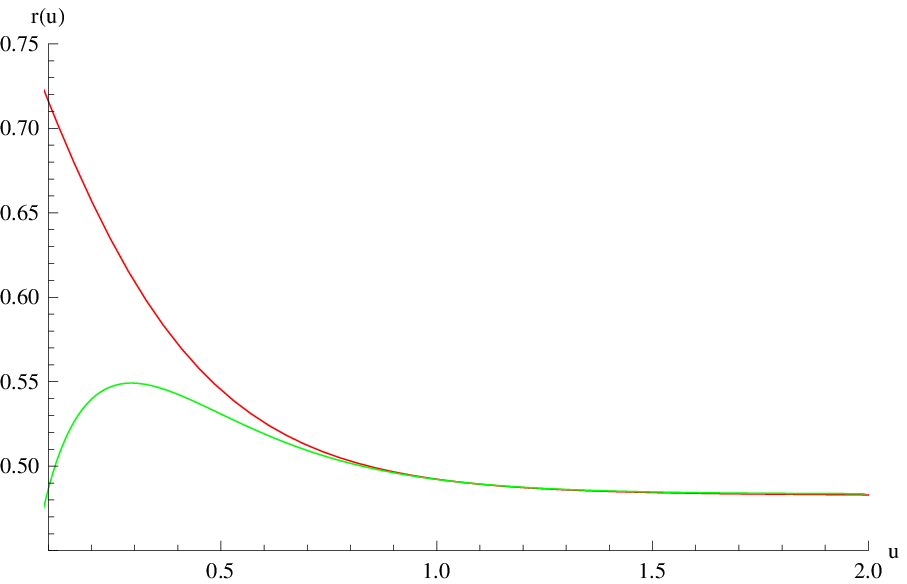}
\includegraphics[width=200pt]{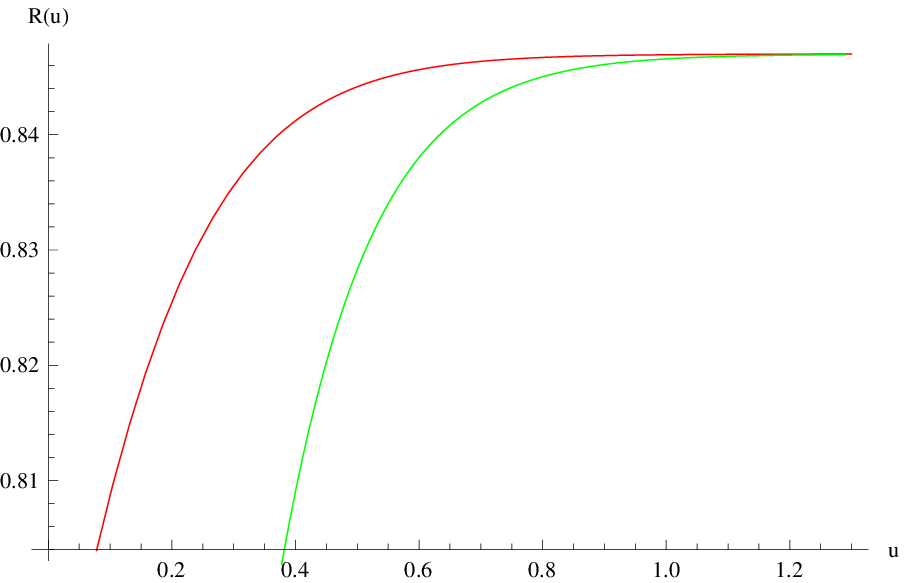}
\end{center}
\caption{{\it The dynamical evolution of apparent horizon (green) and local event horizon (red) for planar black holes $k=0$ in $r$ (left) and $R$ (right) coordinate, respectively. We have set $g^2\,,q$ to unity and $\alpha=2$. }}
\label{fig3}\end{figure}
In Fig.\ref{fig3}, we plot the apparent horizon and the local event horizon as a function of the advanced time for planar black holes with $k=0$. We find that the
apparent horizon is always enclosed by the local event horizon in the dynamical process. Note that in $r$ coordinate, both horizons become decreasing functions of
the advanced time at late times. However, this does not mean the black objects are radiating energies. To clarify this point, we also plot the horizons in luminosity coordinate. We see that both horizons grow with the advanced time monotonically and approach the event horizon at the future infinity. For sphrical/hyperbolic black holes, we also find qualitatively similar features.
\begin{figure}[ht]
\begin{center}
\includegraphics[width=200pt]{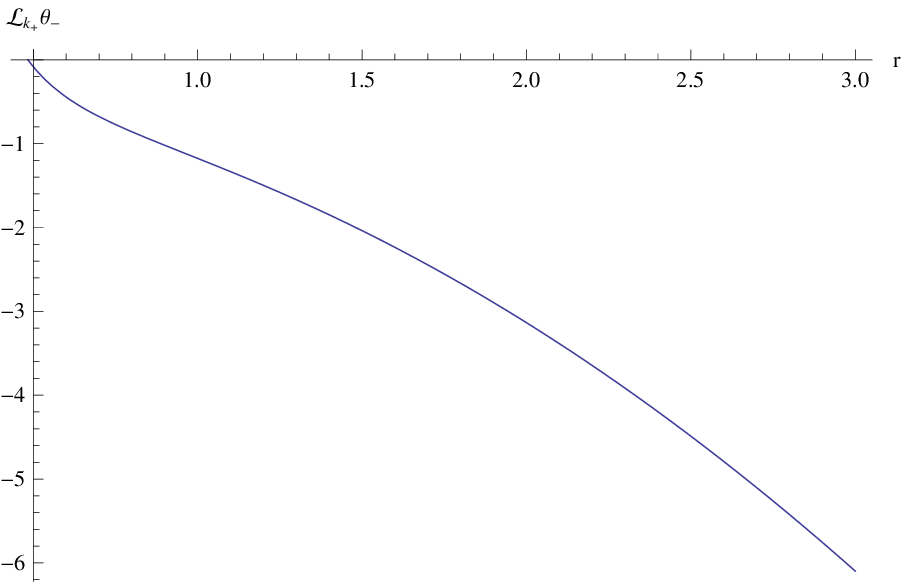}
\includegraphics[width=200pt]{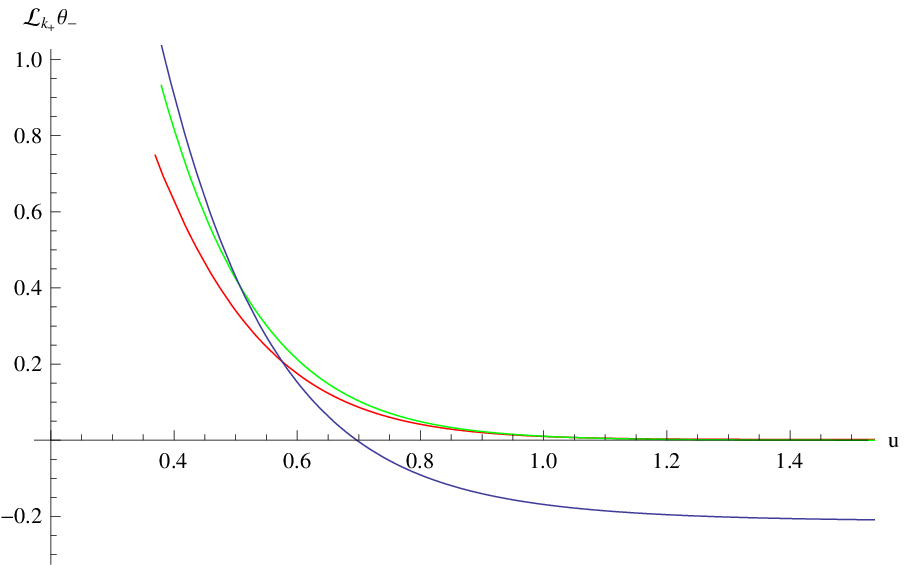}
\end{center}
\caption{{\it The plots of the Lie derivative $\mathcal{L}_{k_+}\theta_{-}$ for planar black holes. The left plot is presented for static solutions and the Lie derivative is plotted as a function of $r$. In the right plot, the Lie derivative is plotted as a function of the advanced time $u$ for $r=r_{AH}$ (Green), $r=r_{EH}$ (Red) and $r=0.53$ (Blue), respectively. We have set $g^2\,,q$ to unity and $\alpha=2$.  }}
\label{fig4}\end{figure}

In addition, the Vaidya mass of the dynamical black holes is given by
\be \label{timemass}
M(u)= \ft{1}{2}M_0\tanh(\ft12\alpha^2 q\,u)\big
(3-\tanh^2(\ft12\alpha^2 q\,u)\big)\ge 0\,,
\ee
which is a monotonically increasing function of the advanced time and approaches the final black hole mass $M_0=\ft{1}{12}\alpha q^3$. This is independent of the radial coordinates.

At the future infinity, the system exponentially reaches the static limit. We find
\be a=q\Big(1-2e^{-\alpha q u}+2 e^{-2\alpha q u}+\cdots \Big)\,,\qquad M=M_0\Big(1-6e^{-2\alpha q u}+16e^{-3\alpha q u}+\cdots  \Big) \,.\ee
The characteristic relaxation time is given by $u_0=1/\big(\alpha q \big)$, which is inversely proportional to the cube root of the final black hole
mass $u_0\sim 1/M_0^{1/3}$.

Finally, we plot the Lie derivative $\mathcal{L}_{k_+}\theta_{-}$ for planar black holes in Fig.\ref{fig4}. It is clear that for static solutions the Lie derivative vanishes on the event horizon and becomes negative outside the event horizon. In dynamical process, it smoothly decreases and approaches the equilibrium value at late advanced times.
Hence, our dynamical solutions have future outer apparent horizons at the future infinity which implies that the solutions indeed describe black holes formation.

\section{More charged black holes}
 In order to construct further solutions in the non-minimal theory Eq.(\ref{gen-nonmin-lag}), we introduce an additional Maxwell field which is non-minimally coupled to the scalar filed. Its Lagrangian density is given by
\be \mathcal{L}_A=-\ft 14 Z^{-1}F^2 \,,\ee
where $F=dA$ and $Z=Z(\phi)$ is the gauge coupling function which is specified by
\be Z=\fft{\gamma\big((n-3)\phi^2+n-1 \big)\phi^{\fft{2}{n-2}}}{(1-\phi^2)^3} \,, \label{zphi}\ee
where $\gamma$ is a positive constant (recall that $\phi<1$). From Maxwell equations
\be \triangledown_{\mu}\big( Z^{-1}F^{\mu\nu}\big)=0\,, \ee
the gauge field strength can be solved as
\be F=\fft{Q\,Z}{\kappa_0\,r^{n-2}}dr\wedge du \,.\ee
Here $Q$ is the total electric charge
\be Q\equiv \kappa_0 \int_{\Sigma_{n-2}} Z^{-1}{}^*F \,.\ee
Provided the gauge coupling function Eq.(\ref{zphi}) and the scalar potential Eq.(\ref{nonpotential}), our new theory admits charged dynamical solutions
\bea ds^2&=&-fdu^2+2dudr+r^2 dx^idx^i \,,\qquad \phi=\Big(\fft{a}{r}\Big)^{\fft12(n-2)}\,,\nn\\
        f&=&g^2 r^2 - \ft{\alpha a^{n-1}}{r^{n-3}}\, {}_2F_1[1,\ft{n-1}{n-2};\ft{2n-3}{n-2};
\ft{a^{n-2}}{r^{n-2}}]\nn\\
      &&+\ft{256(n-1)^3\gamma Q^2}{(n-2)^5 a^{n-3}r^{n-3}}{}_2F_1[1,\ft{n-1}{n-2},\ft{1}{n-2};\ft{a^{n-2}}{r^{n-2}}]\,,\eea
where $a(u)$ satisfies a new non-linear second order differential equation
\be
\fft{\ddot a}{a^2} - \fft{2\dot a^2}{a^3} + \fft{\tilde{\alpha}\dot a}{a}+ \fft{\tilde{\gamma} Q^2\dot{a}}{a^{2n-3}}=0\,,
\qquad \tilde{\gamma}=\fft{128(n-3)(n-1)^3\gamma}{(n-2)^5} \,.
\label{aeqcharge}\ee
We see that both the metric function $f$ and the dynamical evolution equation of the ``scalar charge" receive new contributions from the electric charge.
From the asymptotical behavior of the metric function $f$ when $r\rightarrow \infty$
\be f=g^2 r^2-\ft{1}{r^{n-3}}\big(\alpha a^{n-1}-\ft{256(n-1)^3\gamma Q^2}{(n-2)^5 a^{n-3}}\big)+\cdots \,,\ee
we can read off the new Vaidya mass
\be M=\fft{(n-2)^2}{128(n-1)\pi}\big(\alpha a^{n-1}-\ft{256\gamma(n-1)^3Q^2}{(n-2)^5 a^{n-3}} \big)\,.\label{newbhmass}\ee
The positiveness of the Vaidya mass requires
\be a(u)>\big(\ft{256(n-1)^3 \gamma Q^2}{\alpha(n-2)^5}\big)^{\ft{1}{2n-4}}=\big(\ft{(n-1)\tilde{\gamma}Q^2}{(n-3)\tilde{\alpha}}\big)^{\ft{1}{2n-4}}>0\,,\label{abound}\ee
which provides a lower bound for the ``scalar charge". We shall require the initial state of the evolution satisfying this bound. In fact, in order to ensure the existence of an apparent horizon, the ``scalar charge" should be sufficiently
 large such that the $\alpha$ term dominates in the metric function $f$. This is expected since an apparent horizon exists for generic $\alpha$ and
any non-vanishing $a$ in the neutral limit. Generally speaking,
$a(u)$ satisfies this bound when an apparent horizon exists in the initial state.

The dynamical evolution equation Eq.(\ref{aeqcharge}) can also be integrated, giving rise to a first order equation
\be
\dot a + \tilde\alpha\, a^2\log \big(\fft{a}{q_*}\big)-\ft{\tilde{\gamma} Q^2}{2(n-2)}a^{6-2n}=0\,,\label{firstorder2}
\ee
where $q_*$ is an integration constant. It follows that for $a\leq q_*$, $\dot a>0$, indicating that $a=q_*$ is not a stable point. In fact, we find that there is
only one stable point (denoted by $a=q$) in this equation, which is determined by
\be x \log x=\ft{\tilde{\gamma}Q^2}{\tilde{\alpha}}q_*^{4-2n}\,,\qquad x\equiv \big(\fft{q}{q_*}\big)^{2n-4}\,,\ee
 or more explicitly
\be q_*=q\,\mathrm{exp}\Big(-\ft{\tilde{\gamma}Q^2}{2(n-2)\tilde{\alpha}}q^{4-2n} \Big) \,,\ee
which turns out to be smaller than $q$. Moreover, for $q_*\gg 1$ we find
\be q\simeq q_*(1+\ft{\tilde{\gamma}Q^2}{2(n-2)\tilde{\alpha}}q_*^{4-2n}) \,,\ee
which is a little bigger than $q_*$. We shall point out that this stable point corresponds to the final stable black hole state.

\begin{figure}[ht]
\begin{center}
\includegraphics[width=230pt]{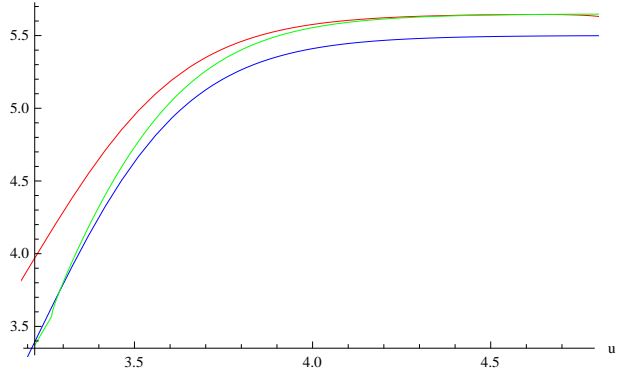}
\end{center}
\caption{{\it The plots for $a(u)$ (the blue line), the apparent horizon (the green line) and the local event horizon (the red line) in the dynamic process.
We have set $g^2\,,\tilde{\alpha}\,,\tilde{\gamma}$ to unity and $q_*=5\,,Q=1/\sqrt{3}$.}}
\label{fig6}\end{figure}
In general, Eq.(\ref{firstorder2}) cannot be solved analytically. We will study it using numerical approach. Without loss of generality, we focus on the $n=4$ dimension.
In Fig.\ref{fig6}, we plot $a(u)$ as a function of the advanced time $u$. It is clear that the space-time evolves from some unstable states in which the power-law
singularity at the origin was dressed by an apparent horizon. This is characteristic for our charged solutions. At the future infinity, the space-time approaches
the equilibrium and arrives at stationary black hole states. We also see that both the apparent horizon and the local event horizon monotonically increase with the advanced time and approach the event horizon of the final
 black holes  at the future infinity.

In addition, the Vaidya mass satisfies
\be \dot M=\ft{(n-2)^2}{64(n-1)\pi}a^{n-2}(\tilde{\alpha}+\tilde{\gamma}Q^2a^{4-2n})\dot a \,.\ee
It follows that $\dot M>0$ in the whole dynamic process since $\dot a>0$, implying that the Vaidya mass also monotonically increases with the advanced time.

At the future infinity, the system approaches the static limit at an exponential rate of $e^{-u/u_0}$. We find
\bea &&a=q\big(1-c_0 e^{-\fft{u}{u_0}}+\ft{3-(2n-7)\log x}{2(1+\log x)}c_0^2 e^{-\fft{2u}{u_0}}+\cdots\big)\,,  \nn\\
      &&M=M_0\Big(1-\ft{(n-3)(n-1)(1+\log x)}{n-3-(n-1)\log x}c_0 e^{-\fft{u}{u_0}}+\ft{(n-3)(n-1)\big(n+1-3(n-3)\log x\big)}{2\big(n-3-(n-1)\log x \big)} c_0^2 e^{-\fft{2u}{u_0}}+\cdots \Big)\,.\eea

\begin{figure}[ht]
\begin{center}
\includegraphics[width=190pt]{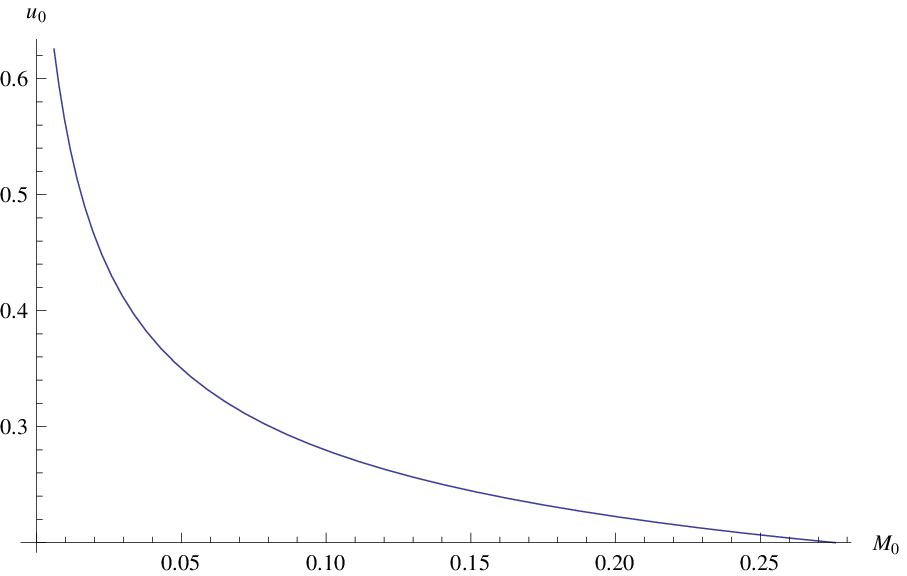}
\includegraphics[width=190pt]{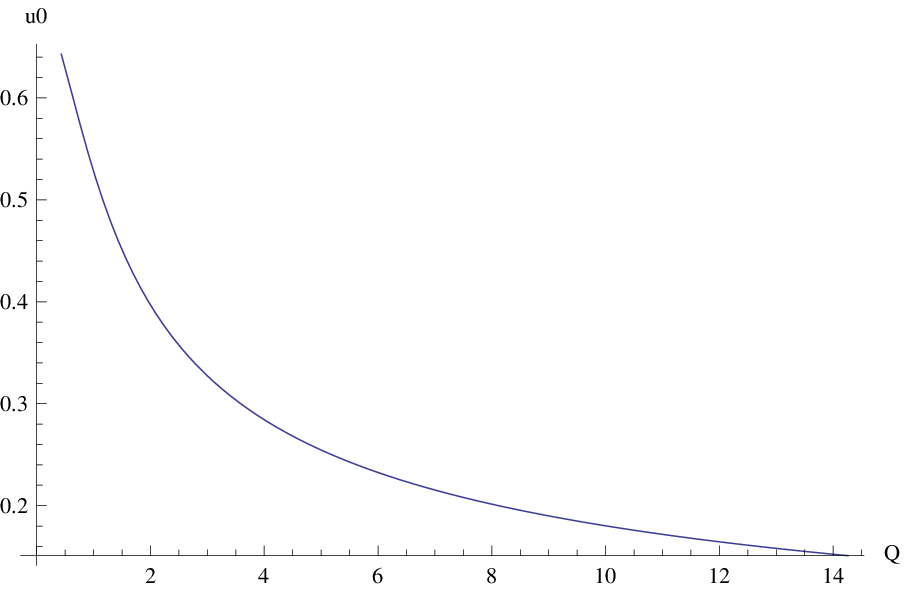}
\end{center}
\caption{{\it The plots for the relation time $u_0$ in the $n=4$ dimension. In the left plot, $u_0$ is plotted as a function of the final black hole mass $M_0$ for
fixing electric charge $Q=1/\sqrt{3}$. In the right plot, it is plotted as a function of the electric charge for fixing black hole mass $M_0=1/(48\pi)$.
We have set $g^2\,,\tilde{\alpha}\,,\tilde{\gamma}$ to unity.}}
\label{u0}\end{figure}
Here $c_0$ is a positive integration constant and $M_0$ is the mass of the final black holes, given by Eq.(\ref{newbhmass}) with $a$ replaced by $q$.
Note that the sub-leading correction of the Vaidya mass requires $\log x<\ft{n-3}{n-1}$. This is equivalent to the condition for the positiveness of the
final black hole mass.
The relaxation time is given by
\be u_0=\fft{1}{\tilde{\alpha}q (1+\log x)}=\fft{1}{q\big(\tilde{\alpha}+\tilde{\gamma}Q^2q^{4-2n}\big)} \,,\ee
which turns out to be a rather involved and decreasing function of the mass and the electric charge of the final black holes
(see Fig.\ref{u0}).


\section{Conclusions}

In this paper, we study the global properties of some exact dynamical black holes with scalar hair which were found in certain minimally and non-minimally coupled
Einstein-scalar gravity. In order to analyze the global structures, we adopt an effective notion ``local event horizon" as well as the apparent horizon.

For the solutions of both minimal and non-minimal thoeries, we find that the apparent horizon is always inside the local event horizon in the whole dynamical process and
both of them increase monotonically with the advanced time and
approach the event horizon at the future infinity. We also find that the apparent horizon smoothly and gradually evolves into a future outer one at the late times of
the evolution. This is instructive to support the solutions describing black holes formation.
At the future infinity, the solution reaches
the static limit exponentially fast with the relaxation time inversely proportional to the $1/(n-1)$ power of the final black hole mass.

There exists a crucial difference in the initial state of the solutions between the minimal and non-minimal theories.
For the non-minimal case, the dynamical evolution of the space-time starts from pure AdS vacua which is linearly stable.
However, driven by some non-linear effects the space-time spontaneously evolves into stationary black hole states at the future infinity.
For the minimal case, the space-time evolves from some finite advanced time (which was set to zero), at which the scalar vanishes. However, the space-time does not become AdS vacua at this point.
It contains a
power-law singularity at the origin which results to linear instability and triggers the dynamical evolution.

For the non-minimal theory, we also obtain new charged solutions by introducing an additional Maxwell filed which is non-minimally coupled to the scalar.
The initial state of the evolution  has non-vanishing scalar and becomes unstable at the linear level. This is significantly different from the pure neutral solutions.
At the future infinity, the charged solutions also exponentially approach stable black hole states with the
relaxation time strongly modified by the electric charge.

\section*{Acknowledgments}
Z.Y. Fan and B. Chen are supported in part by NSFC Grants No.~11275010, No.~11335012 and No.~11325522. H. Lu is supported in part by NSFC grants NO. 11175269, NO. 11475024 and
NO. 11235003.

\end{document}